# Phase-coherent asynchronous optical sampling system

**Honglei Yang,[1,2,*] Shengkang Zhang,[1] Huan Zhao,[1] and Jun Ge[1]**

[1] *Science and Technology on Metrology and Calibration Laboratory, Beijing Institute of Radio Metrology and Measurement, Beijing 100854, China*
[2] *State Key Laboratory of Precision Measurement Technology & Instruments, Department of Precision Instrument, Tsinghua University, Beijing 100084, China*
**yhlpc@163.com*

**Abstract:** Mutual phase coherence is the most crucial factor in asynchronous optical sampling system, and its enhancement is commonly achieved by phase-locking both femtosecond lasers to a shared narrow-linewidth cavity-stabilized laser. Here we report such a system with a low residual optical phase jitter at a level of 0.04 rad in a Fourier frequency band from 1 Hz to 5 MHz around 1.55 μm, corresponding to a timing jitter of 30 as. The residual phase jitter reaches 1 rad at a Fourier frequency between 0.06 Hz and 0.1 Hz, from which the phase-coherence time is inferred at least 10 s. The multi-heterodyne beats experimentally reveal a hardware-limited phase coherence time of ~8.15 s throughout the direct lasing spectral band.

## 1. Introduction

Asynchronous optical sampling systems established by dual femtosecond lasers possess very attractive features like high temporal resolution, broadband spectral detection and rapid signal acquisition for applications, for instance, ranging, spectroscopy, time distribution [1-7]. Due to slightly detuned repetition rates, the output pairs of optical pulses are sequentially imprinted with femtosecond-resolution linear-scanning time delays, and periodically generate interferometric patterns at the detuning frequency. In the frequency domain, a comb structure of evenly-spaced multi-heterodyne beats is mapped from optical region to radio-frequency band and allows for massive parallel sensing. Nonetheless, the mutual phase coherence is the utmost crucial factor to meet the above requirements of 'linear scanning', 'even spacing', and perfect 'comb structure', which are substantially the requirement of Fourier Transform.

Because of the ultrashort temporal duration, femtosecond pulses provide sharper gate signals compared to picosecond pulses, promoting time interval measurement precision to sub-femtosecond level [1,2]. This kind of time-of-flight method drops the carrier phase during the measurement, and therefore permits a relaxation on mutual coherence. The system could be simply accomplished by dual femtosecond lasers, whose repetition rates are locked to a common radio-frequency standard, leaving their carrier-envelop offset frequencies free-running. Indeed, mutual phase coherence in optical region is low due to the large frequency leveraging factor of femtosecond laser, always ranging from $10^5$ to $10^6$. In several applications, self-seeded difference frequency generation or similar process could avoid the coherence degradation resulted from the free-running carrier-envelop offset frequencies, making the system still works well [3-5]. In order to exploit the carrier phase of femtosecond pulses, highly coherent optical oscillations have to be maintained within one-shot interrogation duration. In this case, optical longitudinal modes of the dual femtosecond lasers are fully stabilized by both optical phase locking and self-referencing. The optical phase locking to common phase-coherent optical frequency reference dramatically decreases the frequency leveraging factor to ~1, keeping the comb modes across the whole spectral bandwidths in phase with an ultra-low phase jitter. As a result, a successive phase-stable interferometric pattern could be observed in a relatively long time duration, and analysis on its carrier-phase enhances time interval measurement precision to attosecond level [6,7]. Benefited from the recent development of laser frequency comb in the past decade, asynchronous optical sampling system could exhibit a more intriguing performance and come into its own as an extremely powerful tool in the field of precision measurements.

Here we present a phase-coherent asynchronous optical sampling system. A narrow-linewidth cavity-stabilized laser is employed as a phase-coherent optical frequency reference to regulate both self-referenced femtosecond lasers. To characterize the phase coherence, we investigate its phase noise spectrum and dispersive interferogram. In addition, we experimentally validate the coherence time by observing the linewidth of single-shot long-term multi-heterodyne beats.

## 2. System architecture

The phase-coherent asynchronous optical sampling system follows our previous work on coherent narrow-linewidth optical frequency synthesis [8]. As presented in Fig. 1, a narrow-linewidth cavity-stabilized laser at 1542.14 nm serves as an optical frequency reference. Since the reference laser and dual femtosecond lasers are settled on separate optical tables, the phase-coherent optical oscillation from the reference laser is separately transferred through ~15 m fiber links to both femtosecond lasers. Along each link, additive phase noise is suppressed by a homebuilt compact all-fiber-device-based fiber noise canceller, which exhibits one order of magnitude improvement than our previous bulky-sized ensembles with free-spaced optics. The additional linewidth broadening and fractional frequency instability are below 1.5 mHz and $4.0\times10^{-17}$ at 1 s, respectively.

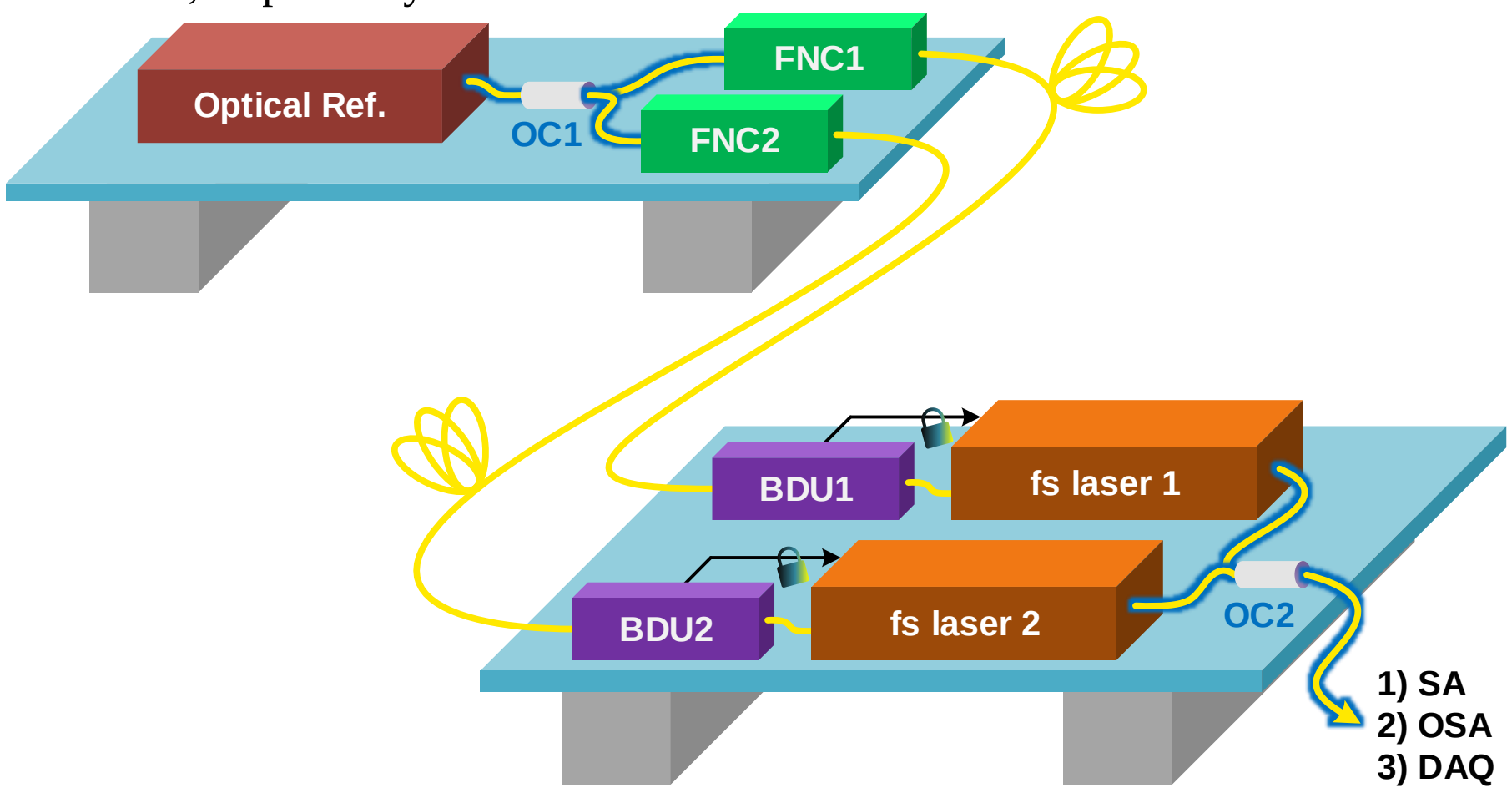

**Fig. 1.** Simplified diagram of the phase-coherent asynchronous optical sampling system. The out-of-loop fibers that could bring in additive phase noise are highlighted in purple and enclosed with sponges firmly fixed on the optical tables. The black arrowed lines denote electronic cables. Optical Ref., optical frequency reference; OC, optical coupler; FNC, fiber noise canceller; BDU, beat detection unit; fs laser, self-referenced femtosecond laser; SA, spectrum analyzer; OSA, optical spectrum analyzer; DAQ, data acquisition board. The yellow curves represent optical fibers.

Two multi-branch Erbium-fiber femtosecond lasers provide broad lasing spectra at ~1550 nm directly, and octave continuums ranging from 1100 nm to 2200 nm via nonlinear optical process. We phase locked the optical beats ($f_{\mathrm{beat},i}$, $i$=1,2) between the transferred optical frequency reference and the adjacent mode of both femtosecond lasers. The carrier-envelop offset frequencies ($f_{\mathrm{ceo},i}$) are stabilized via self-referencing [9]. Subsequently, system characterizations were implemented in the following cases listed in the Table. 1. The phase noise spectra could give quantitative phase coherent time and residual linewidths at separate spectral regions of interest, while the dispersive interferogram exhibits the coherent band and its interferometric contrast. Multi-heterodyne interferogram presents an intuitional view of the residual linewidths at other wavelengths besides those in phase noise characterization, and experimentally validate the first two characterizations. During the measurement, all the phase lock loops are referenced to a hydrogen maser. Since the fibers out of the phase lock loops still bring additive phase noise, they are shorten as possible and enclosed with sponges firmly fixed on the optical tables.

**Table 1. Characterization methods and their corresponding parameter settings**

| Method | Instrument | System parameters | | |
|---|---|---|---|---|
| Phase noise characterization | Spectrum analyzer | $f_{ceo,1}$ - $f_{ceo,2}$ = $f_{beat,1}$ - $f_{beat,2}$ | | $f_{rep,1} = f_{rep,2}$ |
| Dispersive interferometry | Optical spectrum analyzer | $f_{ceo,1} = f_{ceo,2}$, | $f_{beat,1} = f_{beat,2}$, | $f_{rep,1} = f_{rep,2}$ |
| Multi-heterodyne interferometry | Data acquisition board | $f_{ceo,1} = f_{ceo,2}$,* | $f_{beat,1} = -f_{beat,2}$,* | $f_{rep,1} \neq f_{rep,2}$ |

* Not mandatory, but practically adopted.

## 3. Phase noise characterization

In the phase noise characterization, we set the frequency offsets of $f_{beat,i}$ and $f_{ceo,i}$ to be equal, i.e. $f_{ceo,1}$ - $f_{ceo,2}$ = $f_{beat,1}$ - $f_{beat,2}$ = $\Delta f$. With this configuration, the repetition rates are forced to be equal, i.e. $f_{rep,1} = f_{rep,2}$. Pairs of the longitudinal modes of the dual femtosecond lasers beat at an identical frequency of $\Delta f$. Several groups of heterodyne beats at 1530.33, 1542.14, 1550.11, 1563.05, 1564.68 nm with a spectral bandwidth of 0.7 nm are filtered to remove noise from other bands as possible. This results in ~400 pairs of optical longitudinal modes to be simultaneously detected in each passband.

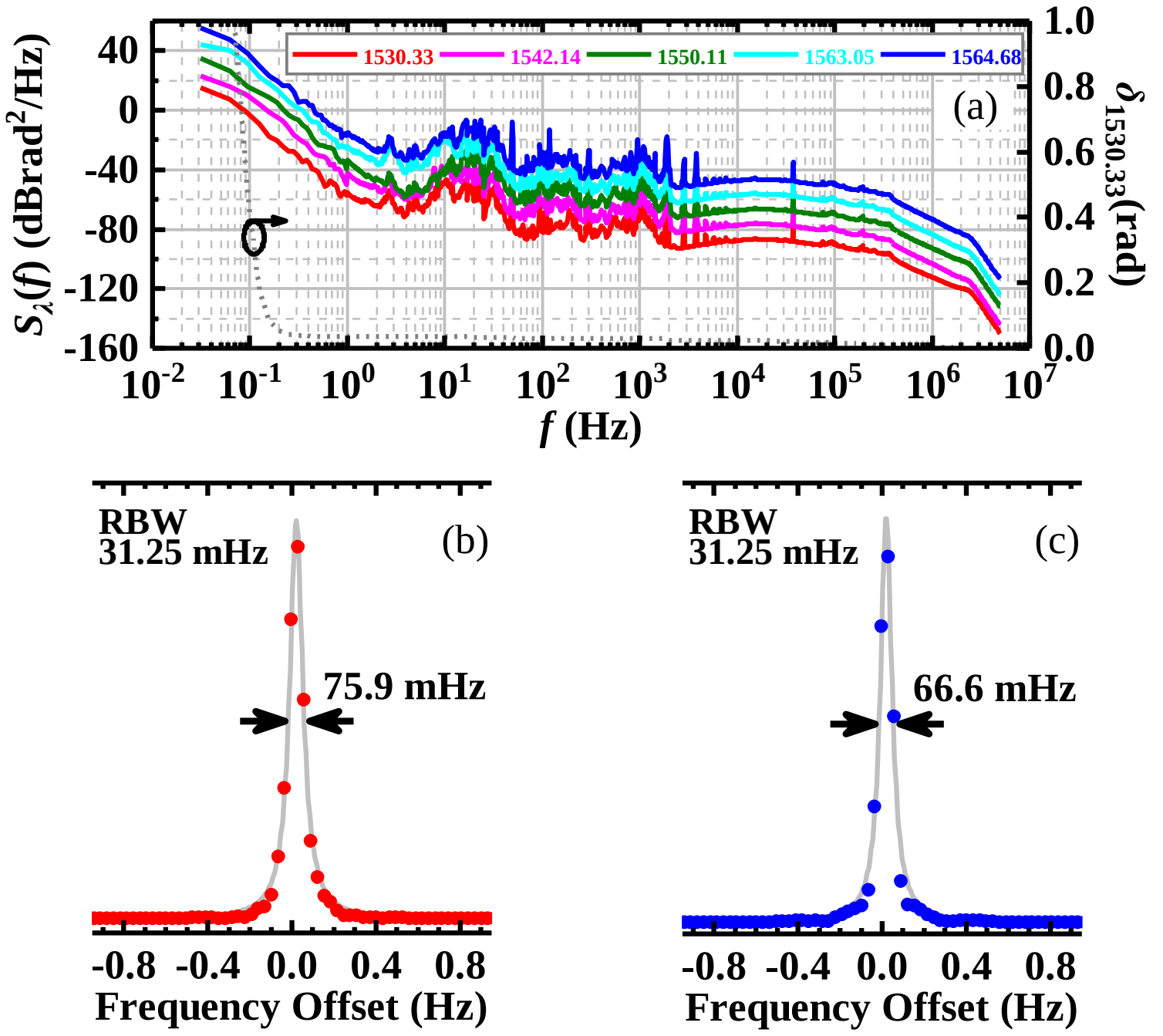

**Fig. 2.** (a) Phase noise power spectrum densities of the asynchronous optical sampling system at five separate spectral regions (left axis) and the integral RMS phase noise at 1530.33 nm (right axis). Expect from the 1530.33-nm curve, the other curves are cumulatively added 10 dB offset for a clear view. (b) and (c) are single-shot RF spectra of optical beats at 1530.33 nm (red dot) and 1564.68 nm (blue dot), respectively, with a resolution bandwidth of 31.25 mHz and their Lorentz fits (gray curve).

Figure 2(a) gives the phase noise power spectrum densities at the five separate spectral regions. The integral RMS phase noise from 5 MHz to 1 Hz at 1530.33 nm is 36.6 mrad, indicating a residual timing jitter approaching to 30 as. This metrics at the other wavelengths maintains this level as well. The coherent peak contains 99.9% of the total power within a 10-MHz bandwidth at 1 s observing time. Then, the integral RMS phase noise continues to go up dramatically and reaches 1 rad at a low Fourier frequency between 0.06 Hz and 0.1 Hz, from which we could estimate a coherent time of at least 10 s. Figure 2(b) illustrates the RF spectra of frequency down-mixed heterodyne beats at 1530.33 nm and 1564.68 nm taken with a resolution bandwidth of 31.25 mHz. The linewidths being in the range from 0.06 Hz to 0.1 Hz validate the phase noise characterization.

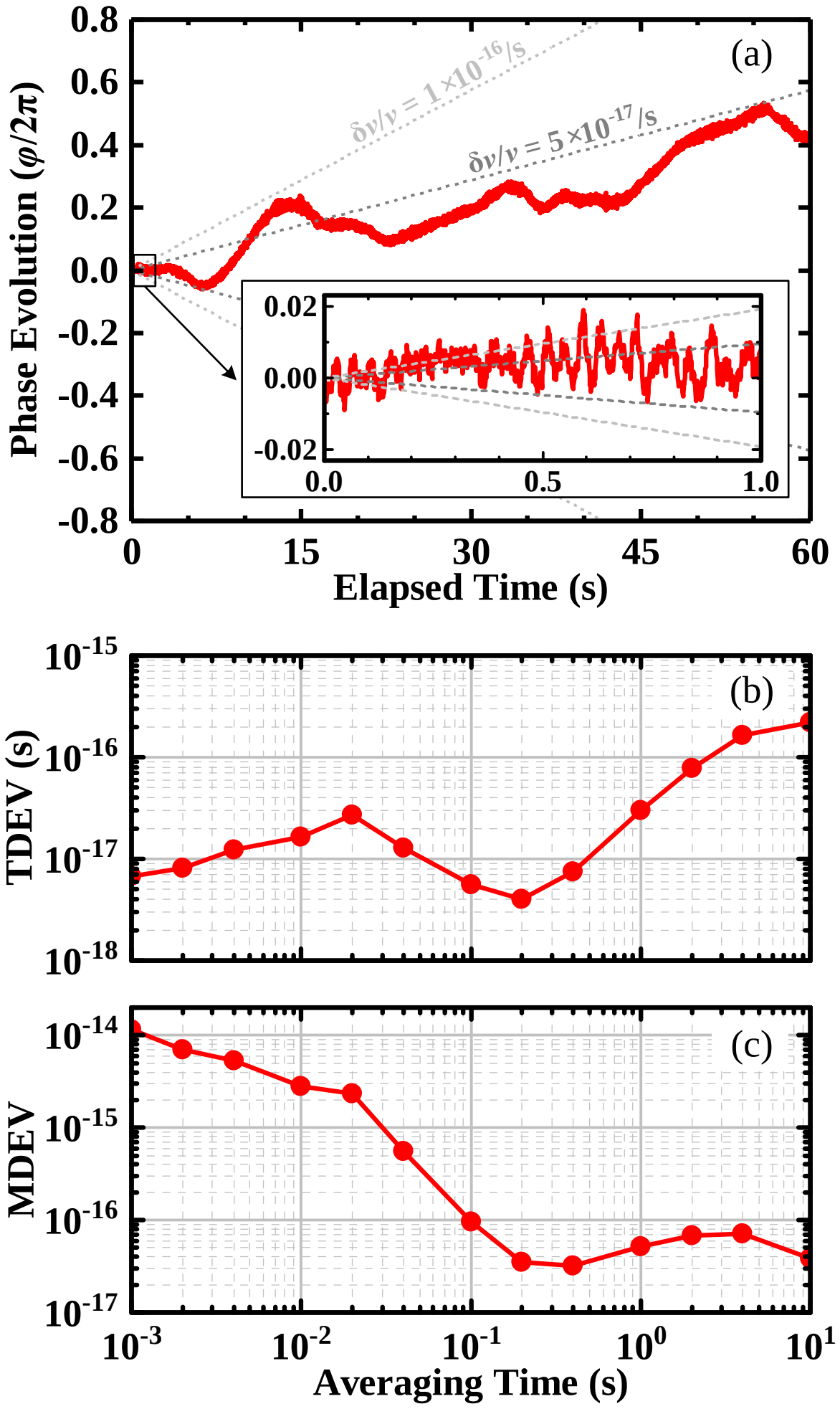

**Fig. 3.** (a) Phase evolution of the grouped modes around 1530.33 nm sampled in a gate time of 1 ms (red curve). Fractional frequency instability limits are indexed in dots. The inset exhibits the zoomed phase evolution in first 1 s. (b) Time deviation (TDEV) for the data in (a). (c) Fractional frequency stability in term of modified Allan deviation (MDEV). MDEV = $(3^{1/2}/\tau)\cdot$TDEV, where $\tau$ is averaging time.

Figure 3(a) gives the phase evolution around 1530.33 nm within 60 s. The phase evolution leads to a frequency instability of less than $1\times10^{-16}$/s in a long timescale. In most time, the system maintains a stability of less than $5\times10^{-17}$/s. We obtained time deviation for the phase evolution data in Fig. 3(b). At an averaging time of 1 s, the time deviation is 30 as, which is exactly same with the integral RMS timing jitter in the above phase noise characterization. In a timescale below 1 s, the system performs a time deviation below 30 as. The bump at averaging times around 0.02 s is mainly caused by mechanical vibration, corresponding to the glitches in the phase noise spectrum at around 500 Hz in Fig. 2(a). While in a timescale above 1 s, the phase drifts due to ambient temperature variation. Figure 3(c) exhibits the fractional frequency stability in term of modified Allan deviation derivative from the time deviation. It is beyond $1\times10^{-16}$/s below an averaging time of 0.1 s, which is inferred by the drastic phase evolution depicted in the inset of Fig. 3(a). At 1 s, the fractional frequency stability is $5.2\times10^{-17}$/s, corresponding to the phase evolution trend in Fig. 3(a).

## 4. Dispersive interferometry

The dispersive interferometry gives an intuitional clue that how broad the phase coherent spectral band is. It is substantially a sort of homodyne interferometry, where the $f_{\text{beat},i}$ and $f_{\text{ceo},i}$ must be identical, i.e. $f_{\text{ceo},1} = f_{\text{ceo},2}$ and $f_{\text{beat},1} = f_{\text{beat},2}$, resulting in identical repetition rates, i.e. $f_{\text{rep},1} = f_{\text{rep},2}$. Since both the femtosecond laser pulse trains are guided via optical fibers, pulse chirp always exists if delicate dispersion compensation has not been done in advance. Next, we briefly introduce the theoretical model of dispersive interferometric pattern generated by chirped pulses. The Gaussian pulses from both lasers are defined as below

$$E_1(t) = E_{01}\exp\left[-(a_1 - ib_1)t^2\right]\exp(i\omega_{c1}t), \tag{2}$$

$$E_2(t) = E_{02}\exp\left[-(a_2 - ib_2)(t-\tau)^2\right]\exp\left[i\omega_{c2}(t-\tau) - i\varphi_0\right], \tag{3}$$

where $E_0$ is electrical field, $a$ is the attenuation factor of Gaussian pulse, $a = 2\ln2/\tau_0^2$, $\tau_0$ is pulse duration, $b$ is chirp rate, $\tau$ is time delay between the pulses, $\omega_c$ is the angular frequency of optical carrier, $\varphi_0$ is a constant phase. By doing Fourier Transform, one can obtain the optical spectra of both pulses.

$$\begin{aligned} E_1(\omega) &= E_{01}\sqrt{\frac{\pi}{a_1 - ib_1}}\exp\left[-\frac{1}{4}\left(\frac{a_1}{a_1^2+b_1^2}\right)(\omega-\omega_{c1})^2\right]\exp\left[-\frac{i}{4}\left(\frac{b_1}{a_1^2+b_1^2}\right)(\omega-\omega_{c1})^2\right] \\ &= \tilde{E}_1(\omega)\exp\left[-\frac{i}{4}\left(\frac{b_1}{a_1^2+b_1^2}\right)(\omega-\omega_{c1})^2\right] \end{aligned} \tag{4}$$

$$E_2(\omega) = \tilde{E}_2(\omega)\exp\left[-\frac{i}{4}\left(\frac{b_2}{a_2^2+b_2^2}\right)(\omega-\omega_{c2})^2\right]\exp\left[i(\omega\tau+\varphi_0)\right] \tag{5}$$

The dispersive interferometric pattern generated by the chirped pulses is written by

$$\begin{aligned} I(\omega) &= \left\langle \left[E_1(\omega)+E_2(\omega)\right]\left[E_1(\omega)+E_2(\omega)\right]^* \right\rangle \\ &= \left|E_1(\omega)\right|^2 + \left|E_2(\omega)\right|^2 + 2\,\mathrm{Re}\left[E_1(\omega)E_2^*(\omega)\right] \\ &\propto \text{const.} + 2\tilde{E}_1(\omega)\tilde{E}_2(\omega)\cos\left\{-\frac{1}{4}\left[\frac{b_1(\omega-\omega_{c1})^2}{a_1^2+b_1^2} + \frac{b_2(\omega-\omega_{c2})^2}{a_2^2+b_2^2}\right] + \omega\tau + \varphi_0\right\} \end{aligned} \tag{6}$$

If unchirped pulses are adopted, i.e. $b_1=b_2=0$, Eq. 6 could be simplified into the common dispersive interferometric pattern.

We recorded the spectral features of both femtosecond lasers plotted in Fig. 4(a), and the dispersive interferometric pattern in Fig. 4(b) with a resolution bandwidth of 0.033 nm in 1 s. Following the guidance in Ref. [10], we adjusted the parameters in Eq. 6 to make the calculation match the measurement. According to the Eq. 2 in Ref. [10], the contrast of the interferometric pattern reaches 99.98% throughout the direct lasing output band, therefore inferring a high phase coherence within the lasing band.

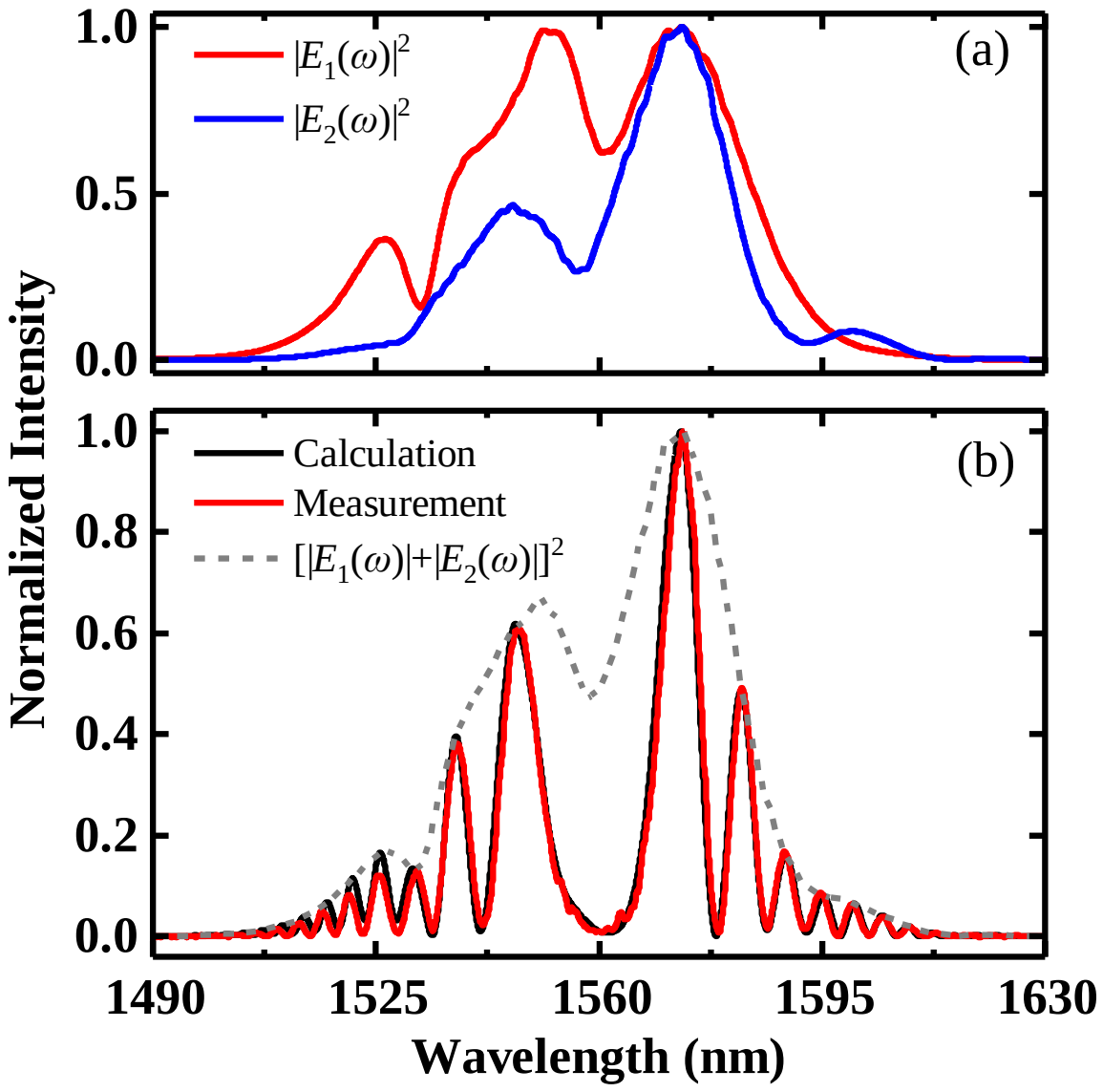

**Fig. 4.** (a) Spectral features of both femtosecond lasers. (b) Calculated and measured dispersive interferometric pattern. The total constructive interferometric pattern is plotted as a reference. All the patterns are recorded at a resolution bandwidth of 0.033 nm. The measurement and calculation are taken within an integration time of 1 s.

## 5. Multi-heterodyne interferometry

In order to evaluate the coherence at other wavelengths besides those in phase noise characterization quantitatively, multi-heterodyne interferometry was done with frequency detuned repetition rates, i.e. $f_{\mathrm{rep},1} \neq f_{\mathrm{rep},2}$, resulted from $f_{\mathrm{ceo},1} = f_{\mathrm{ceo},2}$ and $f_{\mathrm{beat},1} = -f_{\mathrm{beat},2}$. In this case, Pairs of the asynchronous pulses own linearly increasing time delays, and generate pulse-like interferometric peaks periodically with an update rate at the detuned frequency of the repetition rates. Due to a prior expectation about coherence time by the above characterizations, we continuously digitized this periodical interferometric signal for a long time as possible.

After zeropadding the raw data lasting ~8.155 s for a smoother spectral outline, the spectral profile is obtained by Fourier Transform algorithm and presented as Fig. 5(a). The outline agrees with the total constructive interferometric pattern in Fig. 4(b). The multi-heterodyne beats are finely revolved, or so-called comb-tooth resolved, shown as the zoomed spectrum around 1564.68 nm in Fig. 5(b). In Fig. 5(c)-5(f), single comb tooth at 1590.00, 1564.68, 1530.33 and 1515.00 nm are displayed, respectively. Due to the truncation by data acquisition, side lobes are clearly observed around the teeth. Correspondingly, all the teeth are fitted by absolute value function of sinc waveform. Note that even at a spectral regime with a low signal-to-noise ratio, see Fig. 4(f), the fitting functions possessing an identical linewidth match these teeth pretty well. The reason is that the limited sampling duration mainly by the memory of our workstation does not reach the resolution requirement of the temporal coherence. Taking the linewidth broadening coefficient of the truncation window of 1.207 into consideration [11], the measured linewidth in RF domain is 0.123 Hz, which exactly corresponds to the sampling time of according to Nyquist principle. Therefore, the coherence time of at least ~8.15 s is confirmed and agrees with the phase noise analysis. As an expectation, the excellent coherence could maintain across the spectral coverage from 1 μm to 2 μm via nonlinear spectral broadening due to the ultra-low noise feature of the femtosecond lasers [12,13].

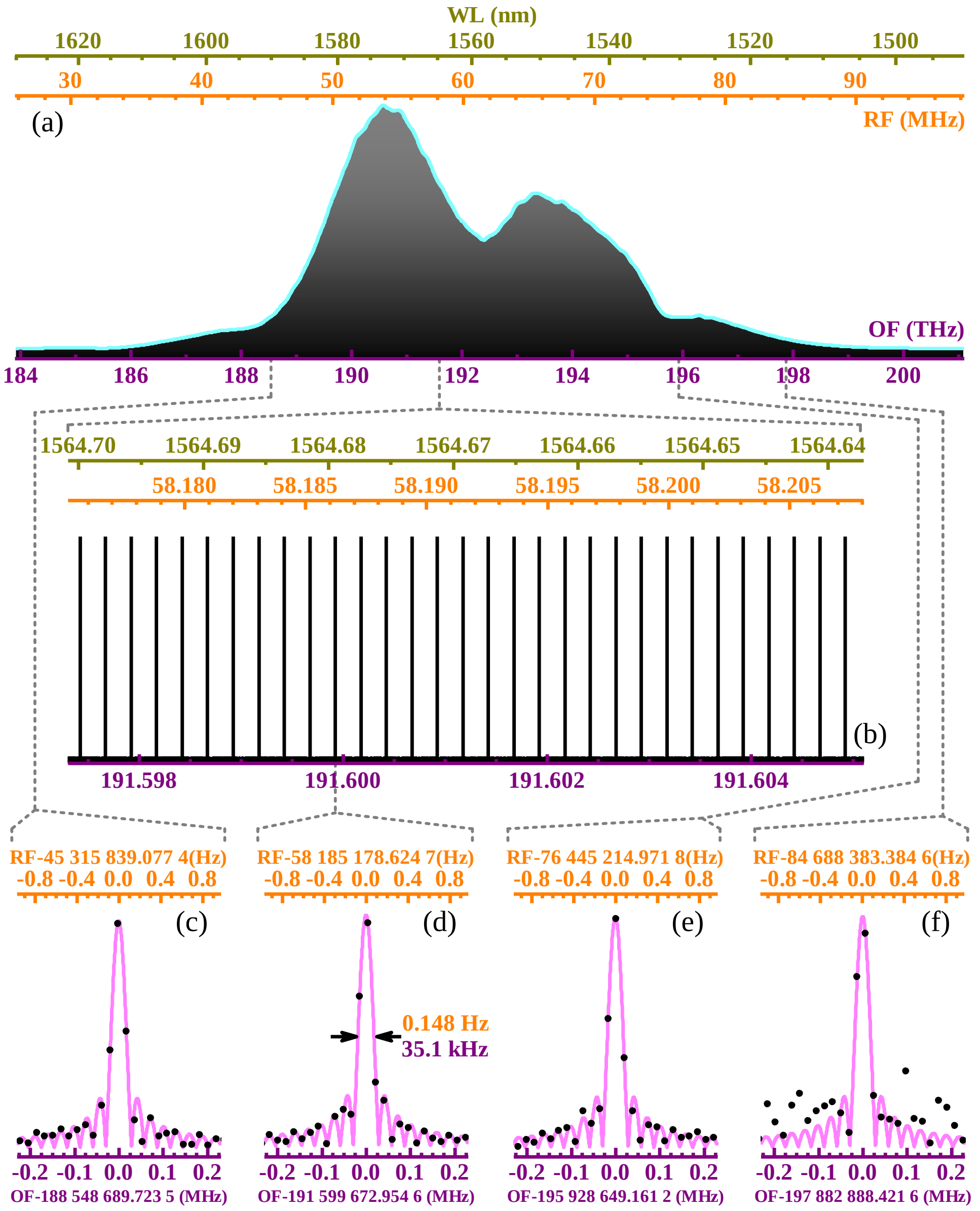

**Fig. 5.** (a) Spectrum of an ~8.155 s long continuously sampled multi-heterodyne interferogram generated by the system. (b) Zoomed spectral band around 1564.68 nm. In (c)-(f), black dots show resolved beat teeth at 1590.00, 1564.68, 1530.33 and 1515.00 nm, respectively. Magenta curves shows their absolute value function fit of sinc waveform. WL, wavelength; RF, radio frequency; OF, optical frequency.

## 6. Conclusion

We have present a phase-coherent asynchronous optical sampling system. The phase noise characterization indicates a residual linewidth of less than 0.1 s, and a corresponding coherence time of at least 10 s. The obtained dispersive interferometric pattern gives a general view of high phase coherence throughout the direct lasing band. The multi-heterodyne spectral analysis gets access to the phase coherence at arbitrary wavelength quantitatively. It infers a hardware-limited phase coherence time of ~8.15 s across the lasing spectral band, further validating the phase noise characterization.

## Funding

This work was supported by the Open Research Fund of the State Key Laboratory of Precision Measurement Technology and Instruments (Grant No. DL18-02).

## Disclosures

The authors declare no conflicts of interest.